# Cooperative Adaptive Cruise Control Design and Implementation

Mustafa Ridvan Cantas, Sukru Yaren Gelbal, Levent Guvenc, Bilin Aksun Guvenc

Automated Driving Lab, Ohio State University

## Abstract

In this manuscript a design and implementation of CACC on an autonomous vehicle platform (2017 Ford Fusion) is presented. The developed CACC controls the intervehicle distance between the target vehicle and ego vehicle using a feedforward PD controller. In this design the feedforward information is the acceleration of the target vehicle which is communicated through Dedicated Short-Range Communication (DSRC) modem. The manuscript explains the detailed architecture of the designed CACC with used hardware and methods for the both simulation and experiments. Also, an approach to overcome detection failures at the curved roads is presented to improve overall quality of the designed CACC system. As a result, the initial simulation and experimental results with the designed CACC system is presented in the paper. The presented results indicate that CACC improves the car following performance of the ego vehicle as compared to the classical Adaptive Cruise Controller.

## Introduction

With the recent advancements in automotive sensors, cars are becoming more autonomous making use of these new technologies. The Advanced Driver Assistant systems such as Adaptive Cruise Control (ACC) system do not only ensure the safety but also increase the comfort of travel. A well-known longitudinal control method, Cooperative Adaptive Cruise Control (CACC), which is an ACC system supported by the Dedicated Short-Range Radio Communication (DSRC) technology that allows Vehicle-to-Vehicle (V2V) communication, enables lower time headway. Reducing the time headway distance between two vehicles can significantly increase the capacity of the road. Also, platooning multiple vehicles has the potential to improve the fuel efficiency of the vehicles by avoiding unnecessary accelerations and decelerations, and reducing the air drag. Motivated by these advantages, in this manuscript, we will present a design and implementation of CACC on an autonomous vehicle platform (2017 Ford Fusion) with initial experimental results.

The designed CACC maintains the desired constant-time headway better than the well-known Adaptive Cruise Control (ACC). Thus, it is possible to reduce the headway for the CACC. In truck platooning smaller time gap results in higher fuel efficiencies by reducing the air drag resistance. Similarly reducing the headway time would increase the capacity of the highways significantly by improving the traffic flow rate [1]. Motivated by these advantages of CACC over ACC, in this manuscript, a Cooperative Adaptive Cruise Controller design process for the autonomous vehicle platform will be explained.

Adaptive Cruise Controllers are already being used in the production vehicles under different names. A comprehensive literature review for ACC systems is done in [2]. Adaptive Cruise Controllers aims to maintain the time-headway constant while car following without breaking the string stability. However, they cannot use low time-headway values since it would result in rear end collision in case of sudden speed changes in the traffic (shock-waves) [3]. Therefore, in ACC a small time gap causes string instability by amplifying the disturbances in the upstream direction. Using the DSRC communication, one can improve the car following performance by reducing the time-headway without breaking the string stability [4]. This car-following model is called Cooperative Adaptive Cruise Control. Some of the earlier work on CACC can be seen in [4- 10]. In [4] authors presented their CACC design methodology by considering the string stability requirements and they experimentally validated their design. One of the early implementations of CACC is done under California PATH program [5-6]. In 2011 several research institutes formed a CACC platoon at Grand Cooperative Driving Challenge (GCDC). Two of the CACC implementations in this challenge can be seen in [7-8]. In [9] authors presented their design for car-following with CACC and approaching maneuver controller. In [10] authors presented multi vehicle look ahead CACC simulation results which shows that the multi vehicle look ahead in CACC improves the performance of CACC. In [11-13] the authors presented how to handle the adversarial environment conditions.

The rest of the paper organized as follows. Next section will explain the designed CACC structure. Following that the simulation environment with target vehicle modeling and simulation results for two vehicle car following scenario will be presented. Then, the experimental vehicle set up with the explanation of sensors will be explained. In the Perception section, in-lane vehicle detection algorithm will be explained. Finally, the manuscript will be concluded with experimental results and their comparison with simulation results.

## CACC Structure

The control structure of the designed CACC system is shown in the block diagram in Figure 1. The designed control system is similar to the one designed and shown to be string -stable in [4]. Since the vehicle does not have built-in ACC the low-level controller is designed as a gain-scheduled PI controller. As an upper controller, a PD controller with a feed-forward controller is used. To sustain the string stability a constant time headway spacing policy is employed [14]. The input of the feedforward controller is the acceleration of the target vehicle which is transmitted through DRSC radio communication. Formulation of the spacing policy is given in Equations 1-2. Where $l$ is the length of the target vehicle, $x_1$ and $x_2$ are the position of the target and ego vehicle, $T_{hw}$ is the desired time-headway and $V_{host}$ is the speed of the host vehicle. The designed PD controller minimizes the spacing error $e$ which is given in Equation 3. Gains of the PD controller chosen as $k_P = k_D{}^2 = w_K{}^2$ where the $w_K$ is chosen close to the bandwidth of the low-level closed-loop bandwidth.

$$\Delta x = x_1 - x_2 - l \qquad (1)$$

$$\Delta x_{desired} = V_{host} T_{hw} + standstill\ distance \qquad (2)$$

$$e = \Delta x - \Delta x_{desired} \qquad (3)$$

Feedforward controller is designed in the same way where it was designed in [8]. Formulation of the feedforward controller is given in Equation 4, where $1/\tau$ represents the desired closed-loop bandwidth.

$$F = \frac{\tau s + 1}{T_{hw} s + 1} \qquad (4)$$

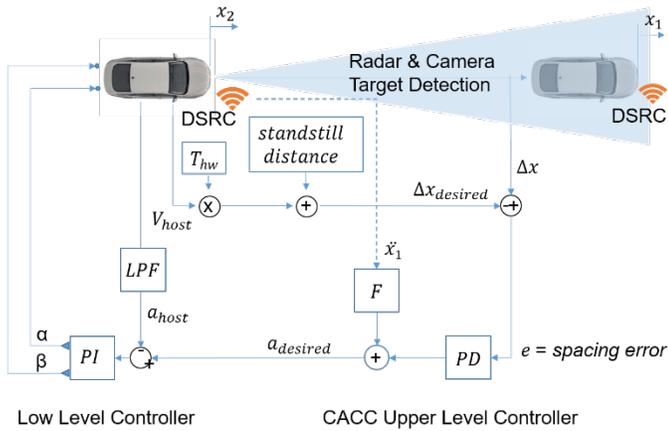

Figure 1. Cooperative Adaptive Cruise Controller (CACC) block diagram.

## Simulation Environment

Development of the initial CACC model is done in CarSim-Matlab co-simulation environment [15]. CarSim is a vehicle simulation environment with the capability of simulating the dynamics of the vehicle. It can also simulate the target vehicle as a kinematic object. In the simulation, target vehicle will be driven by an Intelligent Driver model. By changing the desired speed and/or acceleration limits for the target vehicle, one can create different driving scenarios using Intelligent Driver Model (IDM) [16]. The formulation of the IDM is given in Equations 5-7.

$$\dot{x}_\alpha = \frac{dx_\alpha}{dt} = v_\alpha \qquad (5)$$

$$\dot{v}_\alpha = \frac{dv_\alpha}{dt} = a\left(1 - \left(\frac{v_\alpha}{v_0}\right)^\delta - \left(\frac{s^*(v_\alpha, \Delta v_\alpha)}{s_\alpha}\right)^2\right) \qquad (6)$$

$$s^*(v_\alpha, \Delta v_\alpha) = s_0 + v_\alpha T + \frac{v_\alpha \Delta v_\alpha}{2\sqrt{ab}} \qquad (7)$$

where:
$v_0$: the velocity the vehicle would drive at in free traffic
$s_0$: a minimum desired net distance
$T$: the minimum possible time to the vehicle in front
$a$: the maximum vehicle acceleration
$b$: (a positive number) the maximum vehicle breaking m/s$^2$

The IDM car-following model is commonly used in traffic simulations for simulating driving behavior of the human driver in traffic. In this case, the Intelligent driver model is used to model a human driver for the target vehicle. In CarSim one can also create realistic roads by importing GPS trajectory of the rote. The simulation of the radar and camera is also possible by using the virtual sensors offered in CarSim. Figure 2 shows the visualization of the car following scenario simulation with a radar field of view.

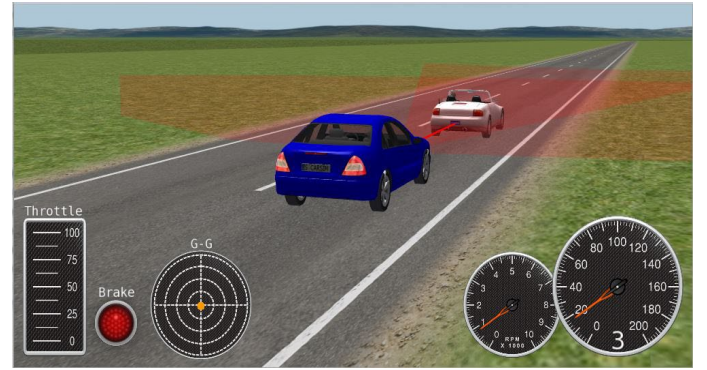

Figure 2. CarSim CACC Simulation visualization.

After creating the simulation environment which replicates the structure shown in Figure 1 simulations run for two different scenarios: ACC and CACC. As the initial evaluation, in the created simulation environment target vehicle first accelerates to a set speed of 20 km/h then it changes set speed to 25km/h and finally it stops. In the simulations, the ego vehicle follows the target vehicle with 1 sec time headway. As it can be seen from simulation results in Figure 3 CACC follows the target vehicle much better. Although both of the speed controllers maintain the time headway, CACC time headway follows the set value (1 second) more accurately. Simulation results with 0.6s is overlayered over the experimental results.

In another scenario, in order to show the performance of the CACC in more realistic scenario the target vehicle speed, acceleration profiles over time is collected by driving the experimental vehicle in an urban route environment. By replaying the recorded data during the simulation, the real world driving experience with a sudden acceleration and braking behavior of the target vehicle in an urban environment is simulated. Similar to the previous simulation results CACC follows the desired time headway of 0.6s much better as compared to ACC (Figure 4). Especially for sudden changes in speed of the target vehicle, CACC responds much better and keeps the desired spacing more accurately.

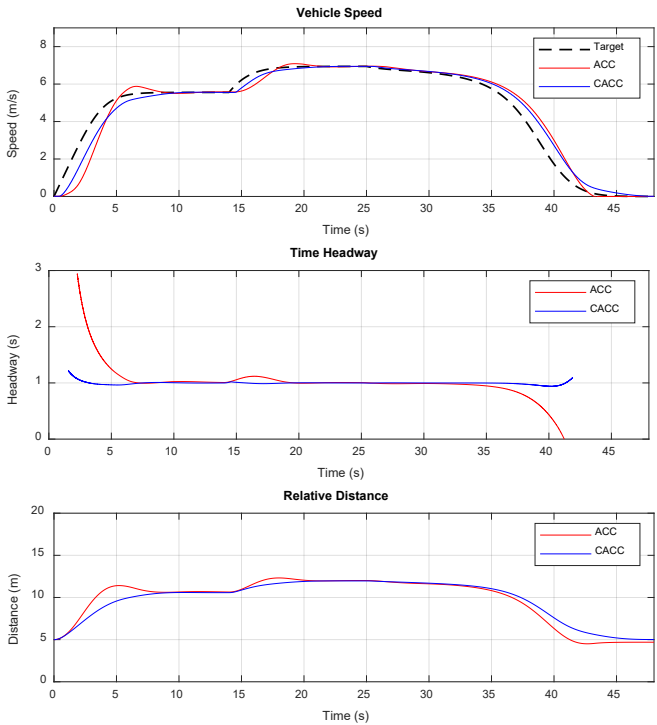

Figure 3. CarSim ACC and CACC simulation results for 1 second time headway with an IDM driven target vehicle.

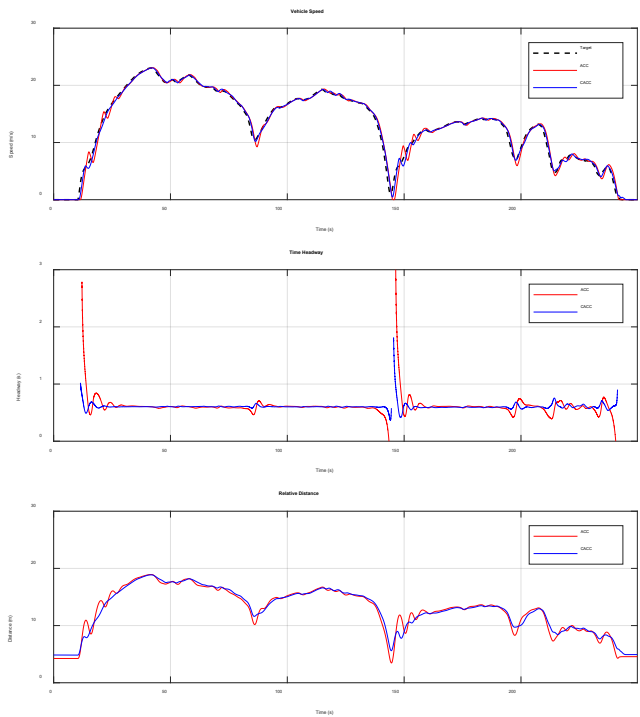

Figure 4. CarSim ACC and CACC simulation results for 0.6 second time headway with a human driven target vehicle data.

# Experimental Vehicle

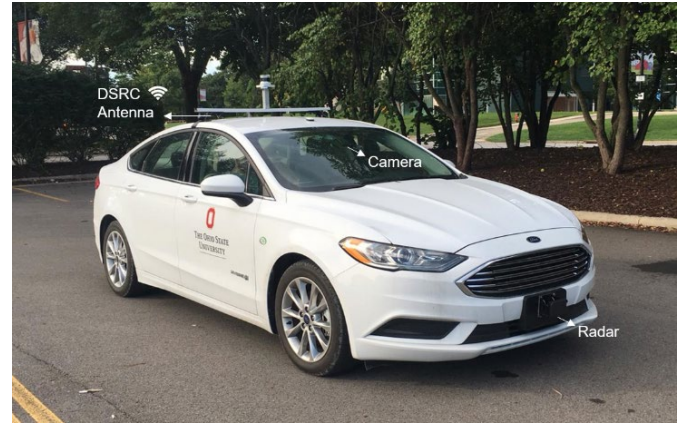

Figure 5. Autonomous vehicle development platform of Automated Driving Lab, at The Ohio State University.

For the experiments, a 2017 Ford Fusion with drive by wire capability is used (Figure 5). Since the vehicle is not equipped with an ACC the throttle and brake actuation are realized through CAN bus messages. For the longitudinal motion measurements, speed and acceleration measurements on the vehicle CAN bus is used.

As the electronic control unit, a dSpace MicroAutoBox II (Figure 6) is employed due to its easy prototyping property with high-performance real-time system implementation capabilities. As it can be seen from the block diagram every equipment in the vehicle is connected to the MicroAutoBox controller. All the measurements coming from the sensors, vehicle CAN bus are processed in the controller and based on the embedded algorithm throttle and brake commands are sent to the vehicle. All the algorithms and the data parsing coming from the sensors are programmed using MATLAB Simulink blocks and they are embedded into the MicroAutoBox controller. As a user interface a portable computer with ControlDesk application is used.

To detect the objects on the road the vehicle is equipped with a 76.5GHz Delphi forward looking radar which can track up to 64 objects and give their positions and relative velocity information. The radar is a combination of both long and middle range radars. Radar is connected to both MicroAutoBox and the laptop. While the data coming from the radar is parsed and processed in the MicroAutoBox, detections can be seen in real time for diagnosis purposes using DataView software. In order to visually validate the radar detections, a forward-looking webcam is connected to the laptop. DataView software can overlay the detections to the video stream acquired from this webcam (Figure 7).

A black and white monocular smart camera from Mobileye (Figure 6) is used to detect lane lines on the road to determine in lane vehicles among detected targets via radar. This camera can detect the lane line markers on the road and provides the lane line information in the form of 3rd order polynomials. Coefficients of the lane line polynomials are available on the CAN bus alongside the road curvature information.

The test vehicle is also equipped with two Denso WSU (Wireless Safety Unit) 5900 DSRC modems to communicate with the target vehicle. In the CACC scenario target vehicle broadcast its acceleration alongside the Basic Safety Message (BSM) [17]. While the first modem is receiving the target vehicle acceleration, the second modem on the vehicle is used to transmit the acceleration of the virtual target vehicle.

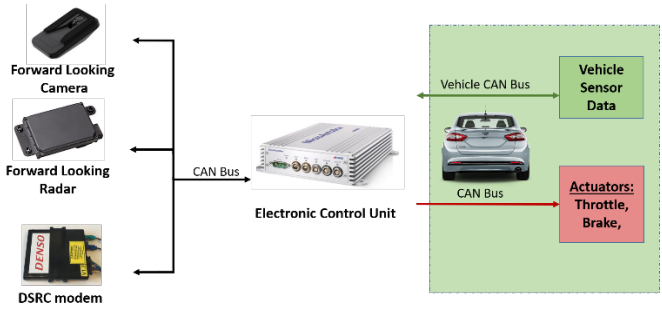

Figure 6. CACC experimental vehicle hardware block diagram.

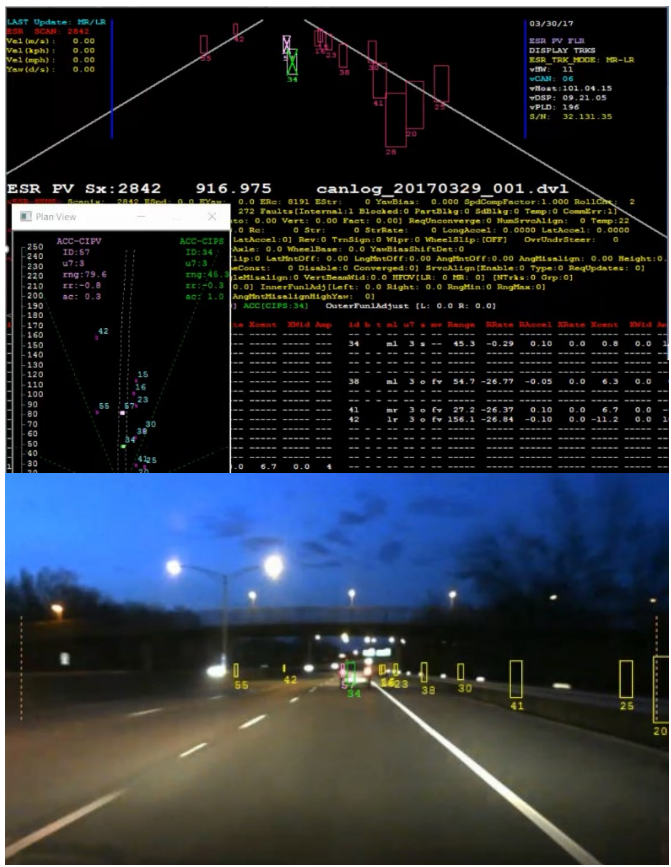

Figure 7. Sample Radar Detection Visualization in real time.

## Perception

A radar is used to detect the positions of the target vehicles. The downside of this method the radar only provides the position and speed information of the detected objects. Since the radar does not know the lane information on the road, it cannot distinguish the vehicles which are in the lane and which are not. From the Radar, an ACC target information is available, which is selected by the radar using the speed of the vehicle, steering angle, and yaw rate information of the ego vehicle. Although most of the time this target detection is valid, since the exact algorithm for choosing the ACC target vehicle is not known and availability of the ACC target depends on the algorithm used in the Radar we are forced to develop another method to detect the vehicles in the lane.

According to Zhang et.al [18], there are two main difficulties to detect the target vehicle using radar. First, differentiation of the lane change or curve entry/exit behavior is challenging. Second, when the vehicle in the next lane goes into the curvature it can be misclassified as in the host lane. To overcome these difficulties a camera and radar are used together. While the lane boundary information comes from the camera, objects detections are acquired from the radar. For each time step, acquired detections are sorted by their longitudinal range. Then each detected object is checked to see whether it is in the host lane or not by comparing its lateral position with respect to the lane boundaries. Block diagram of the system can be seen in Figure 8. One should note that if at least one of the lane lines are not visible to the camera, it is required to create the lane boundaries synthetically. If only one of the lane lines is available, the other lane boundary is created using the available lane boundary information and the lane width. If both of the lane lines are not available, it is assumed that the vehicle moves straight, and the target object is searched within a window where the width of the window is equal to the lane width.

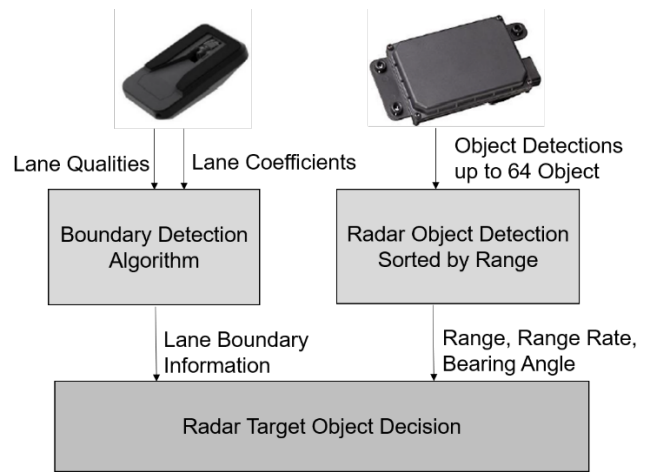

Figure 8. In lane vehicle detection structure with camera and radar combination.

Radar can provide the positions of the detected objects in polar coordinates (Figure 9). The measurements acquired from the radar for each object are listed with their range ($r_i$), bearing angle ($α_i$) parameters and range rate. These coordinates are converted into the Cartesian coordinate system using basic trigonometric equations. Since the radar is placed in front of the front bumper and the origin of the radar coordinate system is chosen as the center of the radar, measurements acquired from the radar are converted to the longitudinal distance ($x_i$) and the lateral distance ($y_i$) of the target objects from the center of the front bumper (Equations 8 and 9).

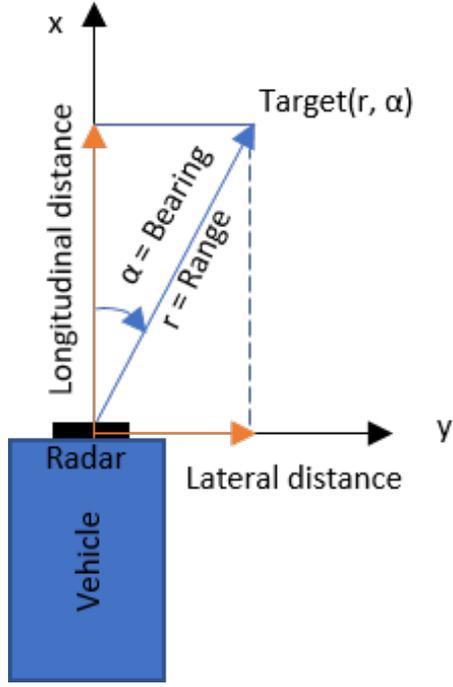

Figure 9. Radar detection coordinate system.

$$Longitudinal\ distance = x_i = r_i * \cos(\alpha_i) \quad (8)$$

$$Lateral\ distance = y_i = r_i * \sin(\alpha_i) \quad (9)$$

The camera in the vehicle is centered at the top of the windshield. It provides the lane line definitions as a third order polynomial in the camera coordinate system, where the origin of the coordinate system is the location of the camera. The curve fitted to the left and right lanes are given in Equations 10 and 11 respectively.

$$y^l(x) = a_0^l + a_1^l * x + a_2^l * x^2 + a_3^l * x^3 \quad (10)$$

$$y^r(x) = a_0^r + a_1^r * x + a_2^r * x^2 + a_3^r * x^3 \quad (11)$$

where $y, x$ represents the lateral and longitudinal positions of the points on the fitted curve. Each $a_i$ represents the coefficients of the fitted curve to the lane lines. Here r and l superscripts differentiate the curve fits for left and right lane respectively. At the final step, by knowing the positions of the targets and lane boundaries, the closest target in the lane can be chosen as the in-lane target. For this purpose, firstly, all the objects are sorted by their longitudinal distances. Then each detected object's coordinates are translated to the camera coordinate system by adding the distance between the camera and the radar in the longitudinal direction ($\Delta x$) Equations 12 and 13.

$$x_{new\_i} = x_i + \Delta x \quad (12)$$

$$y_{new\_i} = y_i \quad (13)$$

Inserting the new lateral distance of the target objects into the Equations 10 and 11, the left (LB) and right boundaries (RB) of the lane at the distance of the target object is calculated (Equations 14 and 15).

$$LB_i = a_0^l + a_1^l * x_{new\_i} + a_2^l * x_{new\_i}^2 + a_3^l * x_{new\_i}^3 \quad (14)$$

$$RB_i = a_0^r + a_1^r * x_{new\_i} + a_2^r * x_{new\_i}^2 + a_3^r * x_{new\_i}^3 \quad (15)$$

If the lateral distance of the $i^{th}$ detected object is between the calculated lane boundaries for the longitudinal distance of the $i^{th}$ object, this object is considered to be an in-lane possible target (Equation 16).

$$(LB_i < y_{new\_i} < RB_i) \quad (16)$$

Among all in-lane possible targets, the closest vehicle in the longitudinal direction is accepted as the in-lane target vehicle. Sample experimental result for this method is presented in Figure 10. The lateral position of the target vehicle with respect to the center of the vehicle and lane boundaries are shown in the plot. One can see from the experimental result that the target vehicle is detected even in the curved section of the road accurately.

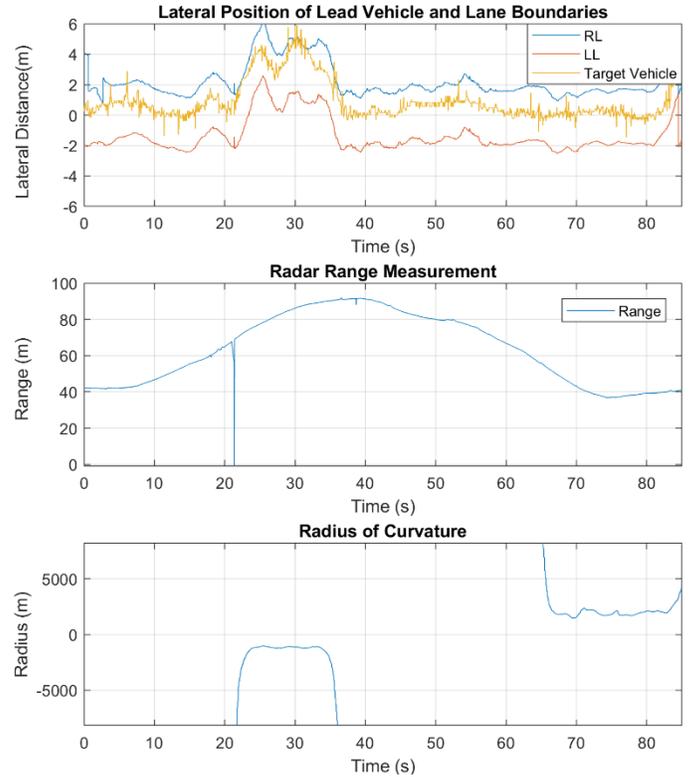

Figure 10. In lane vehicle detection experimental results.

## Experimental Results

In the experiments, the target vehicle speed profile is chosen as the same with the simulation target vehicle speed profile. Target vehicle profile is generated in real time with IDM driver similar to the simulation environment. The virtual target vehicle accelerates to 20 km/h and 25 km/h consecutively then it stops. In the CACC scenario, the simulated acceleration values for the target vehicle is broadcasted through DSRC OBU and it is received by another OBU for the ego vehicle. One can see the experimental results for the ACC and CACC for 1 second headway time overlaid onto the simulation results on Figure 11. The simulation results match with the experimental results. The small mismatches between the experiment and CarSim simulation are caused due to the fact that the CarSim vehicle model is not an exact model of the experimental vehicle. In the CarSim simulation, a generic D class vehicle model is used. In response to the speed changes in the target vehicle speed profile, the CACC speed controllers start accelerating and deceleration faster as compared to ACC using the target vehicle acceleration information coming from the DSRC modem. Thus, the CACC controller can follow the target vehicle more accurately. CACC time headway following performance is much better than the performance of ACC.

Similarly, CarSim simulations and experiments are repeated for desired time headway of 0.6 s. The comparison of the simulation and experiment results are shown in Figure 12. Similar to the previous case the simulation and experimental results are close to each other. And CACC performs better while following the target vehicle with constant 0.6s time headway.

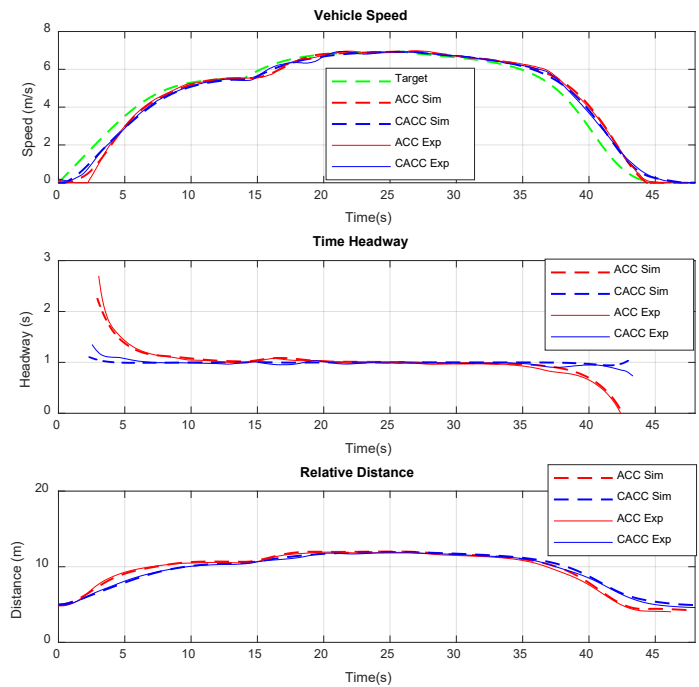

Figure 11. Comparison of ACC and CACC experimental results with simulation results for 1 second desired time headway.

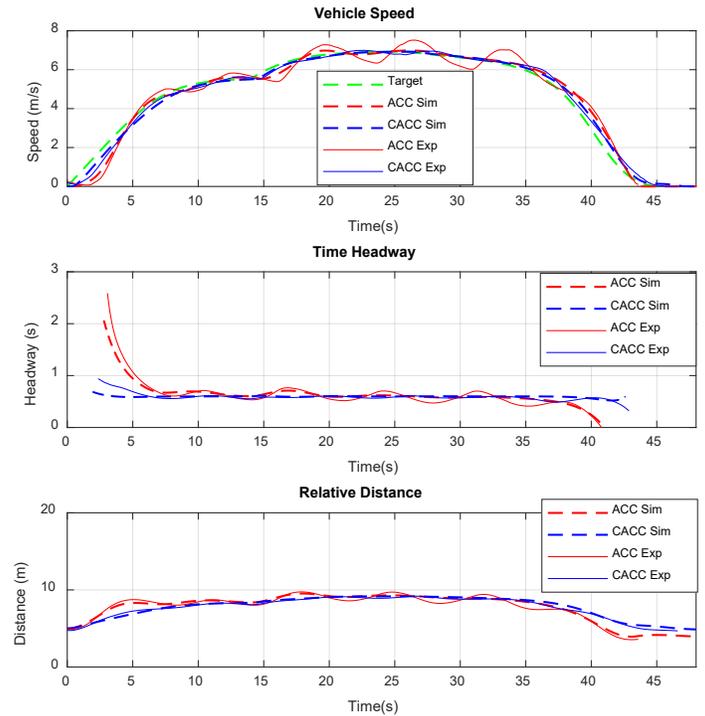

Figure 12. Comparison of ACC and CACC experimental results with simulation results for 0.6 second desired time headway.

## Summary/Conclusions

In this manuscript, the design process of ACC and CACC structure of the Automated Driving Lab at The Ohio State University was presented with initial simulation and experimental results. Note that future work can combine the ACC/CACC approach of this paper with different control, robust control, automotive control, connected and automated driving methods and applications such as those in references [19-89].

Both simulation and experimental results showed that communication between the target vehicle and ego vehicle in CACC increase the car following performance significantly. This performance improvement will lead to better string stability and capacity increase on the roads. As a next step conducted experiments will be repeated with actual target vehicle and higher speed scenarios. The performance of the lower level controllers will be improved. Multi-vehicle look-ahead scenarios will be examined.

## Contact Information

M. R. Cantas is with the Department of Electrical and Computer Engineering, The Ohio State University, Columbus, OH 43210 USA (e-mail: cantas.1@osu.edu).

## Acknowledgments

This work was partially supported the National Science Foundation under Grant 1640308, and by the U.S. Department of Transportation Mobility 21: National University Transportation Center for Improving Mobility (CMU) sub-project titled: Smart Shuttle: Model Based Design and Evaluation of Automated On-Demand Shuttles for Solving the First-Mile and Last-Mile Problem in a Smart City. The authors would like to thank to Nitish Chandramouli and Santhosh Tamilarasan for their valuable discussions and help.

## Definitions/Abbreviations

| | |
|---|---|
| **ACC** | Adaptive Cruise Control |
| **CACC** | Cooperative Adaptive Cruise Control |
| **IDM** | Intelligent Driver Model |
| **BSM** | Basic Safety Message |
| **OBU** | On Board Unit |